\documentclass[12pt]{article}

\makeatletter
\def\eqnarray{%
 \stepcounter{equation}%
 \let\@currentlabel=\theequation
 \global\@eqnswtrue
 \global\@eqcnt\z@
 \tabskip\@centering
 \let\\=\@eqncr
 $$\halign to \displaywidth\bgroup\@eqnsel\hskip\@centering
 $\displaystyle\tabskip\z@{##}$&\global\@eqcnt\@ne
 \hfil$\displaystyle{{}##{}}$\hfil
 &\global\@eqcnt\tw@$\displaystyle\tabskip\z@{##}$\hfil
 \tabskip\@centering&\llap{##}\tabskip\z@\cr}
\makeatother

\newcommand{\Slash}[1]{\ooalign{\hfill/\hfill\crcr$#1$}}

\begin{document}

\title{
Classical Exchange Algebra of the Superstring \\
on $S^5$ with the AdS-Time  }

\author{Shogo Aoyama\thanks{e-mail: spsaoya@ipc.shizuoka.ac.jp} \\
       Department of Physics \\
              Shizuoka University \\
                Ohya 836, Shizuoka  \\
                 Japan}
                 
\maketitle
                 
\begin{abstract}
A classical exchange algebra of the superstring on $S^5$ with the AdS-time is shown on the light-like plane. To this end we use the geometrical method of which consistency is guaranteed by the classical Yang-Baxter equation. The Dirac method does not work,  since there are constraints in which  first-class and second-class constraints are mixed  and one can hardly disentangle with each other keeping the isometry.

\end{abstract}

\vspace{7cm}



\hspace{1.3cm} $AdS_5\times S^5$

\newpage

\section{Introduction}

The AdS/CFT correspondence between the type $IIB$ string theory on $AdS_5\times S^5$ and the $N=4$ supersymmetric QCD is one of the subjects which have been discussed with great interest in the last decade. 
The study of such  string/QCD duality has a long history going back to the late 70's\cite{old}.  The AdS/CFT correspondence\cite{duality} is the most clear-cut assertion 
 of string/QCD duality given ever since.  Correlation functions were extensively studied on both sides of the correspondence to check the duality. Later on 
people discovered a remarkable relationship between the $N=4$ supersymmetric QCD and the Heisenberg spin-chain systems. Namely they claimed that chains of scalar fields of the former theory can be identified with spin-chains of the latter by  calculating the anomalous dimension and the energy for the respective chains\cite{Zarem}. No doubt  integrability of both chain systems is involved  behind this phenomenon. From the  AdS/CFT duality we are then lead to demand a  corresponding integrability on the string side. It is well-known that  the XXX Heisenberg spin-chain  is equivalent to the $O(3)$ non-linear $\sigma$-model at the thermodynamical limit\cite{Faddeev}. Therefore it is quite natural to consider the type $IIB$ string theory on $AdS_5\times S^5$ as a non-linear $\sigma$-model and suspect its integrable structure.  Indeed many works have been done about integrability of non-linear $\sigma$-models on  $S^5$ and $AdS_5$\cite{Zamo, int}.

 Exchange algebrae based on the Yang-Baxter equation are  also characteristic for integrable systems. This aspect of the type $IIB$ string theory on $AdS_5\times S^5$\cite{bei}   attracted  much attention in connection with  higher-loop integrability of $N=4$ supersymmetric QCD\cite{loop}. 
In this paper we will shed a light on the subject from a different angle. 
The following action is a part of 
the type $IIB$ string theory on $AdS_5\times S^5$
\begin{eqnarray}
S=\int d^2\xi \Big[\sum_{a=1}^6 \partial^\mu M^a\partial_\mu M^a -\partial^\mu X^0\partial_\mu X^0\Big],  \label{string}
\end{eqnarray}
where $X^0$ is the AdS-time and $M^a$ are fields constrained on $S^5$  by $\sum_{a=1}^6 M^aM^a=1$. It is supplemented by the Virasoro constraint 
$$
\partial_\pm M^a\partial_\pm M^a=\partial_\pm X^0\partial_\pm X^0.
$$
With a gauge $X^0\propto t$ the action (\ref{string}) becomes the non-linear $\sigma$-model on $S^5$\cite{T}. 
The aim of this paper is to show, though at the classical level, in a rather simple way the exchange algebra  
for the non-linear $\sigma$-model on $S^5$
\begin{eqnarray}
&\quad&\{M^a(\xi^+,\xi^-),M^b(\xi^+,\xi'^-)\}  \nonumber\\
&\quad&\hspace{1cm} =-{1\over 4}\Big[\theta(\xi^--\xi'^-)r^++\theta(\xi'^--\xi^-)r^-\Big]^{ab}_{cd}
M^c(\xi^+,\xi^-)M^d(\xi^+,\xi'^-).    \label{exchange}
\end{eqnarray}
Here $\{\ ,\ \}$ is the Poisson bracket on the light-cone plane $\xi^+=const.$ and $r^{\pm}$ are the classical $r$-matrices given in the fundamental representation of $SO(6)$, which satisfy the classical Yang-Baxter equation.(See eqs. (\ref{r}) and (\ref{YB}).)  As a deformation of (\ref{exchange})  we might think of a quantum exchange algebra 
\begin{eqnarray}
&\quad&M^a(\xi^+,\xi^-)M^b(\xi^+,\xi'^-) \nonumber \\
&\quad&\hspace{1cm}= \Big[\theta(\xi^--\xi'^-)R^++\theta(\xi'^--\xi^-)R^-\Big]^{ab}_{cd} M^d(\xi^+,\xi'^-)M^c(\xi^+,\xi^-),  \nonumber
\end{eqnarray}
by extending the classical $r$-matrices to quantum $R$-matrices as
$$
R^\pm = 1\otimes 1 +\hbar r^\pm + O(\hbar^2).
$$
with $\hbar =-{1\over 4}$. 
The similar classical exchange algebrae in the canonical formalism were discussed  for the 2-dimensional effective gravity in the literature\cite{Ao}. We will proceed the arguments exactly in the same way as for that case.

\section{The Poisson structure}

We shall consider  
non-linear $\sigma$-models on the coset space $G/H$, which are given by the action 
\begin{eqnarray}
S=\int d^2\xi\ {\cal L}={1\over 2}\int d^2\xi\ \eta^{\mu\nu}g_{ij}(X)\partial_\mu X^i\partial_\nu X^j,
  \label{model}
\end{eqnarray}
in  the 2-dimensional flat world-sheet. 
The energy-momentum tensor and the isometry current are respectively  given by 
\begin{eqnarray}
T_{\mu\nu}=g_{ij}(X)\partial_\mu X^i\partial_\nu X^j-\eta_{\mu\nu}{\cal L},
\quad\quad\quad\quad
J^A_{\ \mu}=R^A_{\  i}(X)\partial_\mu X^i.  \nonumber
\end{eqnarray}
Here $R^{Ai}$ are the Killing vectors of the coset space $G/H$, which non-linearly realize the Lie-algebra of $G$ as
\begin{eqnarray}
-R^{Ai}_{\ \ ,j}R^{Bj}+R^{Bi}_{\ \ ,j}R^{Aj}=f^{AB}_{\ \ \ C}R^{Ci},
 \label{Lie} 
\end{eqnarray}
and satisfy the Killing equations
\begin{eqnarray}
 R^A_{\ \{i;j\}}\equiv R^A_{\  i;j} +R^A_{\  j;i}=0. \label{Killing}
\end{eqnarray}
The non-linear $\sigma$-models (\ref{model})  have  conformal invariance and isometry so that 
\begin{eqnarray}
T_{+-}= 0,   \quad\quad
\partial_+T_{--}=0, \quad\quad \partial_-T_{++}=0,   \quad\quad
\partial_+J^A_- &+&\partial_- J^A_+ =0,  \nonumber
\end{eqnarray}
due to the equation of motion
\begin{eqnarray}
\nabla^\mu\partial_\mu X^i \equiv  \partial^\mu\partial_\mu X^i +\Gamma^i_{jk}
\partial^\mu X^j\partial_\mu X^k = 0.
\label{euler}
\end{eqnarray}
We study the canonical structure of the models on the light-like plane setting up the Poisson brackets. 
It can be done by the geometrical method formulated in \cite{Ao}. Namely we may set up the Poisson brackets $\{X^i,X^j\}$ on the light-like plane $\xi^+=const.$ so as to be able to correctly reproduce the diffeomorphism as 
\begin{eqnarray}
\delta_{diff} X^i(\xi^+,\xi^-) &\equiv&\epsilon(\xi^-)\partial_- X^i(\xi^+,\xi^-)  \nonumber\\
          &=& \int d\zeta^- \epsilon(\zeta^-)\{X^i(\xi^+,\xi^-), T_{--}(X(\xi^+,\zeta^-))\},
 \label{diffeo}    
\end{eqnarray}
Here $\epsilon(\xi^-)$ is a local parameter of the transformation. 
It turns out that they are given in the form 
\begin{eqnarray}
  \{X^i(\xi^-),X^j(\eta^-)\}&=&-\{X^j(\eta^-)\,X^i(\xi^-)\} \nonumber \\
& = & -{1\over 4}\theta(\xi^--\eta^-)t^+_{\ AB}R^{Ai}(X(\xi^-))R^{Bj}(X(\eta^-))
\nonumber \\
&\quad &\hspace{1cm}+{1\over 4}\theta(\eta^--\xi^-)t^+_{\ AB}R^{Aj}(X(\eta^-))R^{Bi}(X(\xi^-)).    \label{Poisson}
\end{eqnarray}
The notation of this formula is as follows.  $\theta(\xi^-)$ is the step function and  $R^{Ai}(X(\xi^+))$ are the Killing vectors of the coset space $G/H$, given by (\ref{Lie}) and (\ref{Killing}). The world-sheet coordinate $\xi^+$  in $X^i(\xi^+,\xi^-)$ was omitted to avoid a unnecessary complication.  $t^+_{\ AB}$ is  the most crucial part for our argument in this paper. It is a set of the coefficients taken from the classical $r$-matrices
\begin{eqnarray}
 r^{\pm}=\sum_{\alpha\in R} sgn\ \alpha E_{\alpha}\otimes E_{-\alpha} \pm
   \sum_{A,B} t_{AB}T^A \otimes T^B   
    \equiv  t^\pm_{\ AB}T^A\otimes T^B, \label{r}
\end{eqnarray}
with  $T^A$ the generators of the group $G$ given in the Cartan-Weyl basis as $\{E_{\pm\alpha},H_\mu\}$, $t_{ AB}$ the corresponding Killing metric and 
$sgn\ \alpha=\pm$ according as the roots are positive or negative. The $r$-matrix satisfies the classical Yang-Baxter equation\cite{J}
\begin{eqnarray}
[r_{12},r_{13}]+[r_{12},r_{23}]+[r_{13},r_{23}]=0,  \label{YB}
\end{eqnarray}
 which guarantees the Jacobi identities for the Poisson brackets(\ref{Poisson}), as will be shown later.

First of all we will show that the diffeomorphism (\ref{diffeo}) can be reproduced by using the Poisson brackets (\ref{Poisson}).  
A little algebra yields 
\begin{eqnarray}
\{X^i(\xi^-),T_{--}(X(\zeta^-))\}
 &=& \delta(\xi^--\zeta^-)\partial_- X^j(\zeta^-)g_{jk}(\zeta^-)t_{AB}R^{Ai}(\xi^-)R^{Bk}(\zeta^-)  \nonumber\\
 &-&{1\over 4}\theta(\xi^--\zeta^-)t^+_{\ AB}R^{Ai}(\xi^-)R^B_{\ \{j;k\}}(\zeta^-)\partial_-X^j(\zeta^-)\partial_-X^k(\zeta^-) \nonumber \\
 & +&{1\over 4}\theta(\zeta^--\xi^-)t^+_{\ AB}R^A_{\ \{j;k\}}(\zeta^-)R^{Bi}(\xi^-)\partial_-X^j(\zeta^-)\partial_-X^k(\zeta^-).  \nonumber 
\end{eqnarray}
This becomes 
\begin{eqnarray}
\{X^i(\xi^-),T_{--}(X(\zeta^-))\}=\delta(\xi^--\zeta^-)\partial_- X^i(\zeta^-),
 \label{ddif}
\end{eqnarray}
by the Killing equations (\ref{Killing}) and the property of the Killing vectors\cite{Boul}
\begin{eqnarray}
 t_{AB}R^{Ai}R^{Bj} = g^{ij}.   \nonumber
\end{eqnarray}   
With (\ref{ddif}) the diffeomorphism (\ref{diffeo}) is correctly reproduced.

We remark that the above demonstration would work even if the Poisson brackets ({\ref{Poisson}) were simply given with the Killing metric $t_{AB}$ in place of $t^+_{\ AB}$.
 The specific choice of $t^+_{\ AB}$ given by (\ref{r}) is required to show the Jacobi identities 
\begin{eqnarray}
Q^{ijk}&\equiv& \{X^i(\xi^-),\{X^j(\zeta^-),X^k(\eta^-)\}\}  
+\{X^j(\zeta^-),\{X^k(\eta^-),X^i(\xi^-)\}\}   \nonumber\\
&+&\{X^k(\eta^-),\{X^i(\xi^-),X^j(\zeta^-)\}\} =0.  \label{Jacobi}
\end{eqnarray}
To show it, 
we calculate the quantities $Q^{ijk}$ by (\ref{Poisson}) assuming that $\xi^->\zeta^->\eta^-$. We then use the
 the Lie-Algebra for the Killing vectors (\ref{Lie}) 
to find 
\begin{eqnarray}
Q^{ijk}&=& -{1\over 4}t^+_{\ AB}t^+_{\ CD}\Big[f^{AC}_{\ \ \ E}R^{Ei}(\xi^-)\Big]R^{Dj}(\zeta^-)R^{Bk}(\eta^-)   \nonumber\\
&\quad &-{1\over 4} t^+_{\ AB}t^+_{\ CD}R^{Ci}(\xi^-)\Big[f^{AD}_{\ \ \ E}R^{Ej}(\zeta^-)\Big]R^{Bk}(\eta^-) \label{Q} \\
&\quad &-{1\over 4}t^+_{\ AB}t^+_{\ CD}R^{Ci}(\xi^-)R^{Aj}(\zeta^-)\Big[f^{BD}_{\ \ \ E}R^{Ek}(\eta^-)\Big].   \nonumber
\end{eqnarray}
Note that eq. (\ref{Lie}) can be put in the form
\begin{eqnarray}
R^{Ci}\propto f_{AB}^{\ \ \ C}R^{Ai}_{\ \ ,j}R^{Bj},  \nonumber
\end{eqnarray}
by using $f_{AB}^{\ \ \ C}f^{AB}_{\ \ \ D}\propto \delta^C_D$. Replace all the Killing vectors in (\ref{Q}) by this formula. Then it becomes
\begin{eqnarray}
Q^{ijk}&\propto& {1\over 4}t^+_{\ AB}t^+_{\ CD}\Big(([T^A,T^C])_{EF}\otimes (T^B)_{GH}\otimes (T^D)_{KL}  \nonumber\\
&\quad&\hspace{2cm}+(T^A)_{EF}\otimes ([T^B,T^C])_{GH}\otimes (T^D)_{KL} 
\nonumber \\
&\quad&\hspace{2.5cm}+ (T^A)_{EF}\otimes (T^C)_{GH}\otimes ([T^B,T^D])_{KL}\Big)  \nonumber \\
&\quad&\hspace{1cm}\times R^{Ei}_{\ \ ,l}(\xi^-)R^{Fl} (\xi^-)\cdot R^{Gj}_{\ \ ,m}(\zeta^-)R^{Hm}(\zeta^-)\cdot R^{Kk}_{\ \ \ ,n}(\eta^-)R^{Ln}(\eta^-).   \nonumber
\end{eqnarray}
This  is vanishing due to the classical Yang-Baxter equation (\ref{YB}) so that
  the Jacobi identities (\ref{Jacobi}) are satisfied. 
Thus it has been shown that the Poisson brackets (\ref{Poisson}) are consistent  for the canonical  formalism of the non-linear $\sigma$-models (\ref{model}) on the light-like plane.

\section{Classical exchange algebra}

We will use the Poisson brackets (\ref{Poisson}) to find the classical exchange algebra (\ref{exchange}) in the non-linear $\sigma$-models. Our claim is that such an algebra exists if  the models 
 admit local composite fields $M^a(X), a=1,2\cdots,d$ which change as
\begin{eqnarray}
\delta M^a(X) \equiv \epsilon_A R^{Ai}(X) M^a_{\ ,i}(X)  
        = \epsilon_A (T^A)^a_{\ b}  M^b(X),   \label{compo} 
\end{eqnarray}
 under the isometry transformation $\delta X^i = \epsilon_AR^{Ai}(X)$.  That is, the composite fields $M^a(X)$ belong to a $d$-dimensional representation of the isometry group $G$. 
It depends on the dimension $d$ 
 whether one can construct such composite fields easily or not. We proceed with the argument assuming the existence of them. By using the Poisson brackets (\ref{Poisson}) we find that
\begin{eqnarray}
\{M^a(X(\xi^-)),M^b(X(\zeta^-))\} &=& M^a_{\ ,i}(X(\xi^-))\{X^i(\xi^-),M^b(X(\zeta^-))\}    \nonumber \\ 
&=& M^a_{\ ,i}(X(\xi^-))\{X^i(\xi^-),X^j(\zeta^-)\}M^b_{\ ,j} (X(\zeta^-)) \nonumber \\
&=& -{1\over 4}\theta(\xi^--\zeta^-)t^+_{\ AB}(T^A)^a_{\ c}(T^B)^b_{\ d}M^c(X(\xi^-)))M^d(X(\zeta^-)) \nonumber\\
&+& {1\over 4}\theta(\zeta^--\xi^-)t^+_{\ AB}(T^A)^b_{\ c}(T^B)^a_{\ d}M^c(X(\zeta^-)))M^d(X(\xi^-)).   \nonumber 
\end{eqnarray}
By using the property $t^+_{\ AB}=-t^-_{\ BA}$ and the $r$-matrices (\ref{r}) we get the classical exchange algebra (\ref{exchange}).

We illustrate how to construct the  composite fields $M^a(X)$ for the case of the non-linear $\sigma$-model on 
the coset space $SO(6)/ SO(5)(=S^5)$. The composite fields $M^a(X)$ transforming as (\ref{compo}) in the fundamental representation of $SO(6)$ can be easily constructed. To see this we have recourse to the Callan-Coleman-Wess-Zumino formalism. 
Decompose the generators of $SO(6)$, denoted by $T^A$,  into the $SO(5)$ generators and the coset ones, denoted by $T^{ij}$ and $T^{i6}$ respectively, i.e.,  
$$
\{T^A\}=\{T^{i6},T^{ij}\},\quad\quad i=1,2,\cdots,5
$$
with $T^{ij}=-T^{ji}$ and $T^{i6}=-T^{6i}$. 
Consider a quantity 
$
e^{iX^iT^{i6}}     
$
 with $X^i$ the coordinates parametrizing $S^5$. For an element 
 $g\in e^{i\epsilon_AT^A}$ with real parameters $\epsilon_A$ there exists a compensator $h(X,g)\in SO(5)$ such that
\begin{eqnarray}
ge^{iX^iT^{i6}}= e^{iX^{'i}T^{i6}}h(X,g)     \label{new1}
\end{eqnarray}
This defines the Killing vectors $R^A_i(X)$ as 
\begin{eqnarray}
  \delta X^i=X^{'i}(X)- X^i=\epsilon_AR^{Ai}(X)     \label{R} 
\end{eqnarray}
So far we have not yet specified the representation space of $SO(6)$ for the quantity $e^{iX^iT^{i6}}$. Now we consider it in the fundamental representation as
\begin{eqnarray}
D(\xi)=\left(
\begin{array}{cccc}
\xi^1_{\  1}  &  \cdots & \xi^1_{\  5}    & \xi^1_{\  6} \\ 
\noalign{\vskip0.1cm}
\vdots & \ddots & \vdots  &  \vdots  \\
\noalign{\vskip0.1cm}
 \xi^5_{\  1}  &  \cdots   &  \xi^5_{\  5}  &  \xi^5_{\  6}   \\
\noalign{\vskip0.1cm}
 \xi^6_{\  1}     &\cdots    &  \xi^6_{\  5}     &  \xi^6_{\  6}     \\
\noalign{\vskip0.1cm}
\end{array}
\right)\equiv \left(
\begin{array}{c|c}
 \xi^{i}_{\ j}  &  \xi^{i}_{\ 6} \\
\hline
 \xi^{6}_{\ j}  & \xi^{6}_{\ 6} 
\end{array}\right),\quad\quad \xi\equiv e^{iX^iT^{i6}}    \label{rho}
\end{eqnarray}
The relation (\ref{new1}) implies that 
the last column vector \hspace{-0.3cm} ${\ ^t}(\xi^i_{\ 6}, \xi^6_{\ 6})$ transforms as the fundamental representation of $SO(6)$ by the left multiplication of $g\in e^{iX^iT^{i6}}$, while it does as a singlet of $SO(5)$ by the right multiplication of the compensator $h$. Thus we find  \hspace{-0.3cm} ${\ ^t}(\xi^i_{\ 6}, \xi^6_{\ 6})$ to be the composite fields $M^a(X)$ obeying the transformation law (\ref{compo}) in the fundamental representation. $M^a(X)$ in higher dimensional representations may be considered by making  tensor products of them. 
Take a totally symmetric multiproduct ${\underline 6}\otimes{\underline 6}\otimes\cdots \otimes{\underline 6}$ for instance. Its traceless part gives an irreducible representation of $SO(6)$, say ${\underline d}$, and is decomposed  as 
$
 {\underline d} = {\underline 1}+\cdots
$
under the homogeneous group $SO(5)$. The existence of a singlet implies that, when the quantity $e^{iX^iT^{i6}}$ is written as a $d\times d$ matrix, one of the  column vectors at least, say the last one  \hspace{-0.3cm} ${\ ^t}(\xi^1_{\ d}, \cdots, \xi^d_{\ d})$, can be identified with the composite fields $M^a(X)$ transforming  as (\ref{compo}) in the $d$-dimensional representation. 

 Let us come back to the 6-dimensional $M^a(X)$ obtained above as \hspace{-0.3cm} ${\ ^t}(\xi^i_{\ 6}, \xi^6_{\ 6})$. $D(g)$ is an orthogonal matrix, so that it is  a vector constrained  by $\sum_{a=1}^6 M^aM^a=1$. It denotes a point on $S^5$ parametrized by the polar coordinates as
\begin{eqnarray}
\left(
\begin{array}{c}
M^1\\
 M^2\\
 M^3\\
 M^4\\
 M^5\\
 M^6 
\end{array}\right)
&=&      
\left(
\begin{array}{c}
\cos \Theta^1 \\
\sin \Theta^1\cos \Theta^2 \\
\sin \Theta^1\sin \Theta^2 \cos \Theta^3 \\
\sin \Theta^1\sin \Theta^2 \sin \Theta^3 \cos \Theta^4 \\
\sin \Theta^1\sin \Theta^2 \sin \Theta^3 \sin \Theta^4 \cos \Theta^5 \\
\sin \Theta^1\sin \Theta^2 \sin \Theta^3 \sin \Theta^4 \sin \Theta^5
\end{array}\right)    \label{polar} \\
&\equiv& e^{i\Theta^5 T^{45}}e^{i\Theta^4 T^{34}}e^{i\Theta^3 T^{23}}e^{i\Theta^2 T^{12}}e^{i\Theta^1 T^{61}}
\left(
\begin{array}{c}
 0 \\
 0 \\
 0 \\
 0 \\
 0 \\
 1 
\end{array}\right) \nonumber
\end{eqnarray}
By a left multiplication of $D(g)=D(e^{i\epsilon^AT^A})$ it changes to a vector denoting another point on $S^5$. This induces a transformation of the polar coordinates $(\Theta^1,\cdots,\Theta^5)$ in the fundamental representation as
\begin{eqnarray}
&\ & D(ge^{i\Theta^5 T^{45}}e^{i\Theta^4 T^{34}}e^{i\Theta^3 T^{23}}e^{i\Theta^2 T^{12}}e^{i\Theta^1 T^{61}})    \nonumber\\
&\ & \hspace{3cm}
=D(e^{i\Theta'^{5} T^{45}}e^{i\Theta'^{4} T^{34}}e^{i\Theta'^{3} T^{23}}e^{i\Theta'^{2} T^{12}}e^{i\Theta'^{1} T^{61}}h(\Theta,g))  \label{new2}
\end{eqnarray}
with a compensator such as $D(h(\Theta,g))^i_{\ 6}=\delta_{i6}$\cite{Sa}.
The quantity $e^{i\Theta^5 T^{45}}\hspace{-0.2cm}\cdots\cdots e^{i\Theta^1 T^{61}}$ is not generated by the coset generators alone on the contrary to $e^{iX^iT^{i6}}$ in (\ref{new1}). But both relations (\ref{new1}) and (\ref{new2}) represent a transitive action of $g$ on $S^5$. Therefore (\ref{new2}) defines the Killing vectors for the polar coordinates as 
$$
\delta \Theta^{'i}(\Theta)-\Theta^i=\epsilon^AR^{Ai}(\Theta).
$$
They are different from the ones given by (\ref{R}). We may also set up the Poisson brackets (\ref{Poisson}) with $R^{Ai}(\Theta)$ and  discuss the classical exchange algebra (\ref{exchange}) for $M^a$ given by (\ref{polar}).

\section{Comments}

The reader might think of studying the Poisson brackets (\ref{Poisson}) by using the Dirac method. However we can  see that it does not work. For the model (\ref{model}) we have 
$$
\pi_i \equiv {\delta {\cal L}\over \delta \partial_+ X^i} =g_{ij}\partial_- X^j,
$$ 
which lead us  to a set of constraints 
\begin{eqnarray}
\phi_i(\xi)\equiv \pi_i-g_{ij}\partial_- X^j=0. \label{const}
\end{eqnarray}
They are typical for the Lagrangian which is  of the first degree  in the velocities $\partial_+ X^i$. According to the Dirac method\cite{Dirac} the Hamiltonian
  is merely given by  
\begin{eqnarray}
{\cal H}=\int d\xi^- \lambda^i(\xi)\phi_i(\xi),   \nonumber
\end{eqnarray}
with Lagrangian multipliers $\lambda^i(\xi)$. 
Setting the  Poisson brackets as $\{X^i(\xi^-),\pi_j(\eta^-)\}_P =\delta^i_j\delta(\xi^--\eta^-)$ on the light-like plane $\xi^+=const. $ we find  that
\begin{eqnarray}
C_{ij}(\xi^-,\eta^-)&\equiv& \{ \phi_i(\xi^-), \phi_j(\eta^-) \}_P \nonumber\\ 
&=&-g_{ij}(\xi^-)\partial^\xi_-\delta(\xi^--\eta^-)-g_{ik,j}(\xi^-)\partial_- X^k(\xi^-)\delta(\xi^--\eta^-)  \nonumber\\ 
&\ & +g_{ji}(\eta^-)\partial^\eta_-\delta(\xi^--\eta^-)+g_{jk,i}(\eta^-)\partial_- X^k(\eta^-)\delta(\xi^--\eta^-). 
\label{C}
\end{eqnarray}
Here again the coordinate $\xi^+$  was omitted from $X^i(\xi^+,\xi^-), \pi_i(\xi^+,\xi^-),\phi_i(\xi^+,\xi^-)$ and $g_{ij}(\xi^+,\xi^-)$. They look like second-class constraints. But they contain both first- and second-class constraints as follows. The consistency of the constraints  requires that   
\begin{eqnarray}
\{\phi_i(\xi^-), {\cal H} \}_P =\int d\eta^- \lambda^j(\eta)
\{\phi_i(\xi^-), \phi_j(\eta^-)\}_P =0,     \label{**}
\end{eqnarray}
which  become  $\nabla_-\lambda^i(\xi^+,\xi^-)=0$. With  $\lambda^i=\partial_+X^i$ they are satisfied by means of the equation of motion (\ref{euler}). Therefore the quantity $C_{ij}(\xi^-,\eta^-)$ is not invertible. It means that the constraints (\ref{const}) contain first-class ones. 

A similar problem can be seen in the Green-Schwarz formulation of superparticle or superstring. The Green-Schwarz superparticle\cite{GS} is given by 
$$
S\equiv \int dt {\cal L}={1\over 2}\int dt e^{-1}(\dot X^M -i\bar\Theta\gamma^M\dot\Theta)(\dot X_M -i\bar\Theta\gamma_M\dot\Theta).
$$
The problematic constraint takes the form
\begin{eqnarray}
\phi\equiv \pi_\Theta -i\bar\Theta\Slash{p} =0,  \label{constGS}
\end{eqnarray}
with $\pi_\Theta={\delta {\cal L}\over \delta \dot\Theta}$ and $p_M={\delta {\cal L}\over \delta \dot X^M}$. A simple algebra gives
$$
C_{\alpha\beta}\equiv \{\phi_\alpha, \phi_\beta \}_{+P}=-2i(\gamma^0\Slash{p})_{\alpha\beta}.
$$
 We have  $p^2=0$ as the equation of motion for $e$. Therefore $C_{\alpha\beta}$ is not invertible. Thus the constraints (\ref{constGS}) contain both first- and  second-class constraints. We do not know how to disentangle the two classes of constraints with each other  in a Lorentz-covariant manner. This is why the Green-Schwarz superparticle can not be quantized in a covariant way. 

We do not know either how to do  disentanglement of the constraints (\ref{const})  in an isometrical way. Therefore  the Dirac method  hardly works for the non-linear $\sigma$-models on the light-like plane. 
The author would like to propose that the geometrical arguments in this paper
 give  an alternative way to the Dirac method.  
But note that
\begin{eqnarray}
\int d\eta^- &C_{ij}&(\xi^-, \eta^-)\{X^j(\eta^-),X^k(\zeta^-)\}  \nonumber \\
&=&\delta^k_i\delta(\xi^--\zeta^-)+{1\over 2}\theta(\xi^--\zeta^-)t^+_{AB}R^A_{\ i;j}(\xi^-)\partial_-X^j(\xi^-)R^{Bk}(\zeta^-),    \nonumber
\end{eqnarray}
with the Poisson brackets (\ref{Poisson}) and $C_{ij}(\xi^-, \eta^-)$ given by (\ref{C}). 
  For the special case when  the target space is flat, the second term in the $ r.h.s.$ vanishes due to $R^A_{\ i;j}=0$.  Then
$C_{ij}(\xi^-, \eta^-)$ becomes invertible  as 
\begin{eqnarray}
\Big[C(\xi^-, \eta^-)^{-1}\Big]^{ij}\equiv \{X^i(\xi^-),X^j(\eta^-)\} =
-{1\over 4}\delta^{ij}\Big[\theta(\xi^--\eta^-)-\theta(\eta^--\xi^-)\Big]. \nonumber
\end{eqnarray}
Putting this in other words, eq. (\ref{**}) which becomes $\partial_-\lambda^i=0$ for this case has only a trivial solution  with $\lambda^i$ being an arbitrary function of $x^+$, so that the constraints (\ref{const}) do not contain first-class ones at all. Consequently the Dirac method works in the standard way. 

In this paper we have shown  the exchange algebra of  non-linear $\sigma$-models on the light-like plane $\xi^+=const.$, setting the canonical structure by the geometrical arguments. Of course the exchange algebra of non-linear $\sigma$-models can be discussed also on the space-like plane with $t=const.$. The usual canonical method does work well giving an exchange algebra for the transfer (or monodromy) matrix. But it is not a local field like $M^a$\cite{int}.


\vspace{2cm}

\begin{thebibliography}{99}

\bibitem{old} J. L. Gervais and A. Neveu, ``The quantum dual string wave functional in Yang-Mills theories", Phys. Lett. {\bf B80}(1979)255;

Y. M. Makeenko and A. A. Migdal, ``Exact equation for the loop average in multicolor QCD", Phys. Lett. {\bf B88}(1979)135;

A. M. Polyakov, ``Gauge fields as rings of glue", Nucl. Phys. {\bf B164}(1980)171;

A. A. Migdal, ``QCD=Fermi string theory", Nucl. Phys. {\bf 189}(1981)253.

\bibitem{duality} J. M. Maldacena, ``The large $N$ limit of superconformal field theories and super\-gravity", Adv. Theor. Math. Phys. {\bf 2}(1998)231, hepth/9711200;

S. S. Gubser, I. R. Klebanov and A. M. Polyakov, ``Gauge theory correlators from non-critical string theory", Phys. Lett. {\bf B428}(1998)105, hep-th/9802109;

E. Witten, ``Anti-de Sitter space and holography", Adv. Theor. Math. Phys. {\bf 2}(1998)253, hep-th/9802150.

\bibitem{Zarem} J. A. Minahan and K. Zarembo, ``The Bethe-Ansatz for $N=4$ Super Yang-Mills", JHEP {\bf 0303}(2003)103,  hep-th/0212208;

 N. Beisert, ``The complete one-loop dilatation operator of $N=4$ super Yang-Mills theory", Nucl. Phys. {\bf B676}(2004)3, hep-th/0307015;

 N. Beisert and M. Staudacher, ``The $N=4$ SYM integrable super spin chain", Nucl. Phys. {\bf B670}(2003)439, hep-th/0307042;

G. Arutyunov, S. Frolov, J. Russo and A. A. Tseytlin, `` Spinning strings in $AdS_5 \times S^5$ and integrable systems",  Nucl. Phys. {\bf B671}(2003)3, hep-th/0307191.


\bibitem{Faddeev} L. D. Faddeev, `` How algebraic Bethe ansatz works for integrable model", Les Houches lectures, hep-th/9605187.
 

\bibitem{Zamo} A. B. Zamolodchikov and A. B. Zamolodchikov, ``Relativistic factorized $S$-matrix in two dimensions having $O(N)$ isotopic symmetry", Nucl. Phys. {\bf B133}(1978)525; ``Massless factorized scattering and sigma models with topological termes", Nucl. Phys. {\bf B379}(1992)602;

M. L\"uscher, ``Quantum non-local charges and absence of particle production in two-dimensional non-linear $\sigma$-model", Nucl. Phys. {\bf 135}(1978)1;

D. Bernard, ``An introduction to Yangian symmetries", Int. J. Mod.Phys. {\bf B7}(1993) 3517, hep-th/9211133;

I. Krichever, ``Two-dimensional algebraic-geometrical operators with self-consistent potentials", Func. Anal.  App. {\bf 28}(1994)26;

I. Bena, J. Polchinski, R. Roiban, ``Hidden symmetries of the $AdS_5\times S^5$ superstring", Phys. Rev. {\bf D69}(2004)046002,  hep-th/0305116;

 L. Dolan, C. R. Nappi, E. Witten, ``Yangian symmetry in D=4 superconformal Yang-Mills theory", Contribution to the QTS3 Conference Proceedings, Univ. Cincinnati, September 2003, hep-th/0401243.

\bibitem{int} H. J. de Vega, H. Eichenherr, J. M. Maillet, ``Yang-Bxter algebras of monodromy matrices in integrable quantum field theories", Nucl. Phys. {\bf 240}(1984)377;

N. J. MacKay, ``On the classical origin of Yangian symmetry in integrable field theory", Phys. Lett. {\bf B281}(1992)90;

M. Hatsuda and K. Yoshida, ``Classical integrability and super Yangian of superstring on $AdS_5\times S^5$", Adv. Theor. Math. Phys. {\bf 9}(2005)703, hep-th/0407044;

A. Das, J. Maharana, A. Melikyan and M. Sato, ``The algebra of transition Matrices for the $AdS_5 \times S^5$ Superstring", JHEP {\bf 0412}(2004)055, hep-th/0411200;

M. Forger, M. Bordermann, J. Laarts, U. Sch\"aper, ``The Lie-Poisson structure of integrable classical non-linear sigma models", Commun. Math. Phys. {\bf 152}(1993)167.

\bibitem{bei} C. Gomez and R. Hernandez, ``The magnon kinematics of the AdS/CFT correspondence", JHEP {\bf 0611}(2006)021, hep-th/0608029;

J. Plefka, F. Spill and A. Torielli, ``On Hopf algebra structure of the AdS/CFT $S$-matrix", Phys. Rev. {\bf D74}(2006)066008, hep-th/0608038;



N. Beisert, ``The $S$-matrix of AdS/CFT and Yangian symmetry", nlin.SI/0704.0400;

A. Torielli, ``Classical $r$-matrix of the $su(2|2)$ SYM spin-chain", 
Phys. Rev. {\bf D75}(2007)105020, hep-th/0701281;

S. Moriyama and A. Torielli, ``A Yangian double for the AdS/CFT classical $r$-matrix", JHEP {\bf 0706}(2007)083, hep-th/0706.0884;

N. Beisert and F. Spill, ``The classical $r$-matrix of AdS/CFT and its Lie bialgebra structure", Commun. Math. Phys. {\bf 285}(2009)537, hep-th/0708.1762.

\bibitem{loop} T. Klose and J. Plefka, ``On the integrability of large $N$ plane-wave matrix theory", Nucl. Phys. {\bf B679}(2004)127, hep-th/0310232;

 N. Beisert, ``The $su(2|3)$ dynamic spin chain", Nucl. Phys. {\bf B682}(2004)487, hep-th/0310252;

D. Serban and M. Staudacher, ``Planar $N=4$ gauge theory and the Inozemtsev long range spin chain", JHEP {\bf 0406}(2004)001, hep-th/0401057.


\bibitem{T}  A. A. Tseytlin, ``Spinning strings and AdS/CFT duality", hep-th/0311139.

\bibitem{Ao} F. A. Smirnov and L. A. Takhtajan, ``Towards a quantum Liouville theory with $c>1$", Leningrad preprint;

A. Alekseev and S. Shatashvili, ``From geometric quantization to quantum conformal field theory", Commun. Math. Phys. {\bf 128}(1990)197;

S. Aoyama, ``The classical exchange algebra of the 2$D$ effective supergravity in the geometrical formulation", Phys. Lett. {\bf B256}(1991)416; 
``The classical exchange algebra of the 2$D$ effective $(2,0)$ supergravity in the geometrical formulation", Mod. Phys. Lett. {\bf A6}(1991(2069); 
``Topological gravity with exchange algebra", Phys. Lett. {\bf B324}(1994)303,  hep-th/9311054.

\bibitem{J} M. Jimbo, ``Quantum $R$ matrix related to the generalized Toda system:\ an algebraic approach", Lecture Notes in Physics {\bf 246}(Springer, Berlin ,1986) p. 335.

\bibitem{Boul} D. G. Boulware and L. S. Brown, ``Symmetric space field theory",
 Ann. Phys. {\bf 138}(1982)392.

\bibitem{Sa} A. Salam and J. Strathdee, ``On Kaluza-Klein theory", Ann. Phys. {\bf 141}(1982)317.

\bibitem{Dirac} P. A. M. Dirac, $Lectures\ on\ quantum\  mechanics$, Belfer Graduate School of Science, Yeshiva Univ., New York(1964).

\bibitem{GS} M. B. Green, J. H. Schwarz and E. Witten, $Superstring \ Theory$, Vol. I(Cambridge U. P., Cambridge, 1987).



\end{thebibliography}
\end{document}